\documentclass[twocolumn,preprintnumbers,amsmath,aps]{revtex4}
\usepackage{graphicx}
\usepackage{dcolumn}
\usepackage{bm}
\usepackage{color}
\usepackage{multirow}
\begin{document}
\newcommand{\kvec}{\mbox{{\scriptsize {\bf k}}}}
\def\eq#1{(\ref{#1})}
\def\fig#1{Fig.\hspace{1mm}\ref{#1}}
\def\tab#1{Tab.\hspace{1mm}\ref{#1}}
\title{The isotope effect in ${\rm H_{3}S}$ superconductor} 
\author{R. Szcz{\c{e}}{\'s}niak$^{\left(1, 2\right)}$}
\email{szczesni@wip.pcz.pl}
\author{A. P. Durajski$^{\left(1\right)}$}
\email{adurajski@wip.pcz.pl}
\affiliation{$^1$ Institute of Physics, Cz{\c{e}}stochowa University of Technology, Ave. Armii Krajowej 19, 42-200 Cz{\c{e}}stochowa, Poland}
\affiliation{$^2$ Institute of Physics, Jan D{\l}ugosz University in Cz{\c{e}}stochowa, Ave. Armii Krajowej 13/15, 42-200 Cz{\c{e}}stochowa, Poland}
\date{\today}
\begin{abstract}
The experimental value of ${\rm H_{3}S}$ isotope coefficient decreases from $2.37$ to $0.31$ in the pressure range from $130$ GPa to $200$ GPa. 
We have shown that the value of $0.31$ is correctly reproduced in the framework of the classical Eliashberg approach. On the other hand, 
the anomalously large value of the isotope coefficient ($2.37$) may be associated with the strong renormalization of the normal state by the electron density of states.       
\end{abstract}
\maketitle

\vspace*{0.25 cm}

The metallic hydrogen the most probably could be the superconductor with the very high value of the critical temperature ($T_{C}$) \cite{Wigner1935A}, \cite{Ashcroft1968A}. The expected high $T_{C}$ is associated with the large Debye frequency (the mass of the proton is very small) and the lack of the electrons on the inner shells, which should significantly increase the electron-phonon coupling constant ($\lambda$) \cite{Cudazzo2008A}, \cite{Szczesniak2009A}, \cite{McMahon2011A}. Unfortunately, the pressure of the hydrogen's metallization is very large ($p>400$ GPa \cite{Narayana1998A}, \cite{Stadele2000A}). For this reason, the experimental confirmation of the theoretical predictions has not been obtained to this day.

\begin{figure}
\includegraphics[width=\columnwidth]{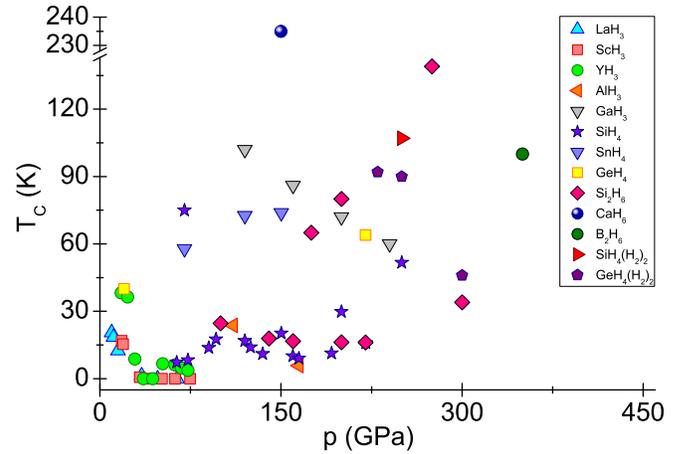}
\caption{The predicted critical temperatures for the hydrogen-rich compounds. The results for tri-hydrides:  
LaH$_3$, ScH$_3$, YH$_3$, AlH$_3$, and GaH$_3$ are from works \cite{Kim2010A}, \cite{Goncharenko2008A}, and \cite{Gao2011A}. 
The results obtained for four-hydrides are presented in the papers: SiH$_4$ \cite{Eremets2008A} (experiment), \cite{Martinez2009A}, \cite{Chen2008A}, SnH$_4$ \cite{Tse2007A}, GeH$_4$ \cite{Gao2008A}. The results for six-hydrides and eight-hydrides were obtained in works Si$_2$H$_6$ \cite{Jin2010A}, \cite{Flores2012A}, $\rm Ba_{2}H_{6}$ \cite{Abe2011A}, $\rm CaH_{6}$ \cite{Wang2012A}, SiH$_4$(H$_2$)$_2$ \cite{Li2010A}, and GeH$_4$(H$_2$)$_2$ \cite{Zhong2012A}, \cite{Zhong2013A}.}
\label{f1}
\end{figure}

In 2004 Ashcroft suggested the existence of the superconducting state in the hydrogen-rich compounds with the critical temperature comparable to $T_{C}$ of the pure hydrogen, whereas the metallization pressure might be subjected to the significant decrease due to the existence of the chemical pre-compression \cite{Ashcroft2004A}. Ashcroft's predictions were confirmed in many later papers. The selected results are presented 
in \fig{f1}.

\begin{figure}
\includegraphics[width=\columnwidth]{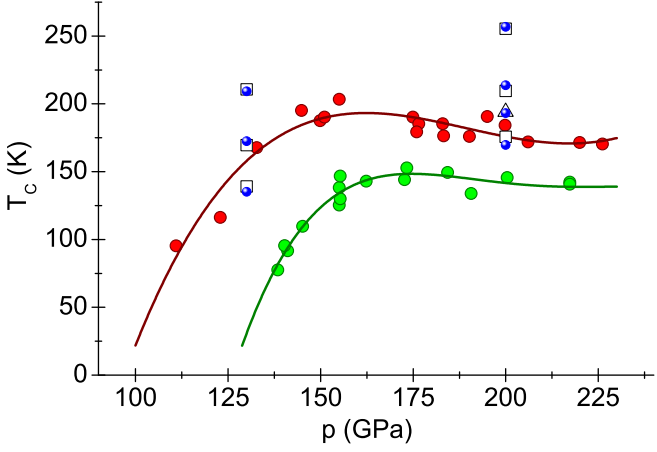}
\caption{The influence of the pressure on the value of the critical temperature - ${\rm H_{3}S}$ (the red circles) and ${\rm D_{3}S}$ 
(the green circles) \cite{Einaga2015A}. The lines were obtained using the approximation procedure. The squares represent the results obtained with the help of the classical Eliashberg equations in the harmonic approximation, the triangle represents the anharmonic analysis, and the blue spheres denote expression \eq{r5}.}
\label{f2}
\end{figure}

The superconducting state in the hydrogen sulfide with the exceptionally high value of the critical temperature ($T_{C}\sim 200$ K) was discovered in 2014 \cite{Drozdov2014A}, \cite{Drozdov2015A}. The detailed dependence of the critical temperature on the pressure for the compounds ${\rm H_{3}S}$ and ${\rm D_{3}S}$ is presented in \fig{f2}. 

The experimental results \cite{Drozdov2014A}, \cite{Drozdov2015A} and the theoretical papers \cite{Duan2014A, Li2014A, Durajski2015A, Durajski2015B, BianconiEPL, Cappelluti} suggest that the superconducting state in the hydrogen sulfide is induced by the electron-phonon interaction. In particular, the strong isotope effect was observed. However, the values of the isotope coefficient ($\alpha$) significantly 
differ from the canonical value of $0.5$ predicted by the BCS theory \cite{Bardeen1957A}, \cite{Bardeen1957B}. 

In the presented paper, we explained the experimental data for $\alpha$ on the basis of the classical and the extended Eliashberg formalism basing on the phonon pairing mechanism.

\begin{figure}
\includegraphics[width=\columnwidth]{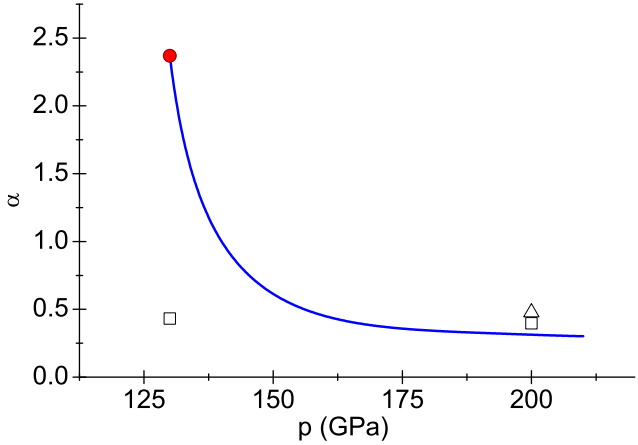}
\caption{The blue line - the experimental values of the isotope coefficient on the basis of formula \eq{r1}. The squares were obtained in the framework of the classical Eliashberg formalism in the harmonic approximation. The triangle corresponds to the classical Eliashberg formalism - the anharmonic analysis. The red circle was obtained assuming the strong renormalization of the normal state by the electron density of states.}
\label{f3}
\end{figure}

In the first step, on the basis of the experimental results, we determined the approximation lines $T^{\rm H_{3}S}_{C}\left(p\right)$ and $T^{\rm D_{3}S}_{C}\left(p\right)$ which served for the calculation of the isotope coefficient:

\begin{equation}
\label{r1}
\alpha_{\rm exp}\left(p\right)=-\frac{\ln\left[T^{\rm D_{3}S}_{C}\left(p\right)\right]-\ln\left[T^{\rm H_{3}S}_{C}\left(p\right)\right]}
{\ln\left[m_{\rm D}\right]-\ln\left[m_{\rm H}\right]},
\end{equation}
where $m_{\rm D}$ and $m_{\rm H}$ are respectively the deuterium's and protium's atomic mass. The shape of the function $\alpha_{\rm exp}\left(p\right)$ is plotted in \fig{f3}. It can be clearly seen that the isotope coefficient decreases with the increasing pressure. In particular, the following values were obtained: $\alpha_{\rm exp}\left(130 \hspace{1mm}{\rm GPa}\right)=2.37$ and $\alpha_{\rm exp}\left(200 \hspace{1mm}{\rm GPa}\right)=0.31$.

The value of the isotope coefficient for $p=200$ GPa can be reproduced in the framework of the classical Eliashberg formalism. To this end, we solved numerically equations \cite{Eliashberg1960A}, \cite{Szczesniak2006A}:
\begin{equation}
\label{r2}
\varphi_{n}=\frac{\pi}{\beta}\sum_{m=-1100}^{1100}
\frac{K\left(i\omega_{n}-i\omega_{m}\right)-\mu^{\star}\left(\omega_m\right)}
{\sqrt{\omega_m^2Z^{2}_{m}+\varphi^{2}_{m}}}\varphi_{m},
\end{equation}
\begin{equation}
\label{r3}
Z_{n}=1+\frac{1}{\omega_{n}}\frac{\pi}{\beta}\sum_{m=-1100}^{1100}
\frac{\lambda\left(i\omega_{n}-i\omega_{m}\right)}{\sqrt{\omega_m^2Z^{2}_{m}+\varphi^{2}_{m}}}
\omega_{m}Z_{m},
\end{equation}
where $\varphi_{n}=\varphi\left(i\omega_{n}\right)$ represents the order parameter function, and $Z_{n}=Z\left(i\omega_{n}\right)$ denotes the wave function renormalization factor. The fermion Matsubara frequency is given by the formula: $\omega_{n}=\frac{\pi}{\beta}\left(2n-1\right)$, 
$\beta=1/k_{B}T$ ($k_{B}$ is the Boltzmann constant). The electron-phonon pairing kernel has the following form: 
$K\left(z\right)=2\int_0^{+\infty}d\Omega\frac{\Omega}{\Omega ^2-z^{2}}\overline\alpha^{2}F\left(\Omega\right)$. The Eliashberg functions 
($\overline\alpha^{2}F\left(\Omega\right)$) for $p=130$ GPa and $p=200$ GPa were calculated by Duan {\it et al.} \cite{Duan2014A}.

The depairing electron correlations in the Eliashberg formalism are described with the use of the formula: 
$\mu^{\star}\left(\omega_{n}\right)=\mu^{\star}\theta \left(\omega_{C}-|\omega_{n}|\right)$. The quantity $\mu^{\star}$ denotes the Coulomb pseudopotential, $\theta$ is the Heaviside function. $\omega_{C}$ represents the cut-off frequency: $\omega_{C}=3\Omega_{\rm{max}}$, where 
$\Omega_{\rm{max}}$ is the Debye frequency. It should be noted that the Coulomb pseudopotential was defined by Morel and Anderson  \cite{Morel1962A}:
\begin{equation}
\label{r4}
\mu^{\star}=\frac{\mu}{1+\mu\ln\left(\frac{\omega_{e}}{\omega_{\rm ln}}\right)}.
\end{equation}
The symbol $\mu$ is given by the formula: $\mu=\rho\left(0\right)U$, whereas $\rho\left(0\right)$ is the value of the electron density of states at the Fermi level, and $U$ is the Coulomb integral. The quantity $\omega_{e}$ represents the characteristic electron frequency and 
the logarithmic phonon frequency is given by:
$\omega_{{\rm ln}}=\exp\left[\frac{2}{\lambda}\int^{\Omega_{\rm{max}}}_{0}d\Omega\frac{\overline\alpha^{2}\left(\Omega\right)F\left(\Omega\right)} {\Omega}\ln\left(\Omega\right)\right]$.

In \fig{f2}, we marked the values of the critical temperature calculated with the help of the Eliashberg equations. 
We considered $\mu^{\star}\in\{0.1,0.2,0.3\}$. Additionally, we also placed the value of $T_{C}$ for $p=200$ GPa, determined beyond the harmonic approximation \cite{Errea2015A}. It turns out that the numerical results can be reproduced using the formula (see also \fig{f1}): 
\begin{equation}
\label{r5}
k_{B}T_{C}=\omega_{{\rm ln}}\exp\left[\frac{-\left(1+\lambda\right)}{\lambda-\mu^{\star}\left(1+0.4369\lambda\right)}\right],
\end{equation}
where the electron-phonon coupling constant should be calculated from: $\lambda=2\int^{\Omega_{\rm{max}}}_{0}d\Omega\frac{\overline\alpha^{2}\left(\Omega\right)F\left(\Omega\right)}{\Omega}$. 

On this basis, it was found out that the values $\mu^{\star}$ corresponding to $\left[T_{C}\right]_{\rm exp}$ were equal to $0.239$ and $0.286$, respectively for the pressure at $130$ GPa and $200$ GPa (the harmonic approximation), and $0.146$ (the anharmonic analysis).

The expression on the isotope coefficient was derived using the dependence:
\begin{equation}
\label{r6}
\alpha=\frac{\omega_{\rm ln}}{2T_{C}}\frac{dT_{C}}{d\omega_{\rm ln}}.
\end{equation}
Thus:
\begin{equation}
\label{r7}
\alpha=\frac{1}{2}\left[1 - 
\frac{\left(1 + \lambda\right)\left(1 +0.4369\lambda\right)\left(\mu^{\star}\right)^2}
{\left(\lambda - \mu^{\star}\left(1 + 0.4369\lambda\right)\right)^{2}}\right].
\end{equation}
The theoretical results have the following form: $\alpha\left(130 \hspace{1mm}{\rm GPa}\right)=0.432$ and $\alpha\left(200 \hspace{1mm}{\rm GPa}\right)=0.397$ (the harmonic approximation), and $\alpha\left(200 \hspace{1mm}{\rm GPa}\right)=0.477$ (the anharmonic approach). It can be easily seen that the theoretical value of the isotope coefficient for $p=200$ GPa in harmonic approximation qualitatively well reproduce the experimental data. In the case of $p=130$ GPa the discrepancy between the Eliashberg result and the result of the measure is extremely high, which means the collapse of the classical theoretical description.

The high value of the isotope coefficient in the terms of the lower pressures can be tried to explain by the pairing mechanism other than the electron-phonon mechanism \cite{Hirsch2015A}. However, the modifying of the classical Eliashberg formalism should also be considered. From the theoretical point of view it highlights the big change of the electron density of states at and near the Fermi surface together with the pressure change. The {\it ab initio} calculations performed for $p=210$ GPa suggest the existence of the sharp peak of $\rho\left(\varepsilon\right)$ very close to the Fermi surface \cite{Papaconstantopoulos2015A}. The peak moves away from the Fermi surface and vanishes for the lower pressures \cite{Duan2014A}, \cite{Li2014A}, \cite{Bianconi2015A}. Hence, physically this means the significant modification of the normal state in the studied system.

Let us consider the renormalized Green function of the normal state, in which the depreciation of the electron density of states was taken into account \cite{Szczesniak2001A}: 
\begin{equation}
\label{r8}
G_{\kvec}\left(i\omega_{n}\right)=
-\frac{i\omega_{n}\tau_{0}+\varepsilon_{\kvec}\tau_{3}}{\omega_{n}^{2}+\varepsilon_{\kvec}^{2}+B^{2}}A
-\frac{i\omega_{n}\tau_{0}+\varepsilon_{\kvec}\tau_{3}}{\omega_{n}^{2}+\varepsilon_{\kvec}^{2}}\left(1-A\right),
\end{equation}
where $\tau_{0}$, $\tau_{3}$ are the Pauli matrices associated with the normal state and $\varepsilon_{\kvec}$ is the electron energy. 
The parameters $A\in\left<0, 1\right>$ and $B$ determine the depth and the width of the decrease in electron density of states with respect to the baseline at the Fermi level. Deriving the Eliashberg equations for the renormalized Green function and using the approximations discussed in paper \cite{Szczesniak2001A}, the algebraic equation on the critical temperature can be obtained: 
\begin{widetext}
\begin{equation}
\label{r9}
1=\frac{\lambda}{1+\lambda}\ln\left(\frac{\omega_{\ln}}{2\pi k_{B}T_{C}}\right)
-\frac{\lambda}{1+\lambda}\left[f_{1}\Psi\left(\frac{1}{2}\right)+
2f_{2}{\rm Re}\Psi\left(\frac{1}{2}+\frac{iB}{2\pi k_{B}T_{C}}\right)+
2f_{3}{\rm Re}\Psi\left(\frac{1}{2}+\frac{igB}{2\pi k_{B}T_{C}}\right)\right],
\end{equation}
\end{widetext}
where:
\begin{equation}
\label{r10}
g=\left[\frac{\left(1-A\right)\left(1+\lambda\right)+A}{1+\lambda}\right]^{1/2},
\end{equation}
\begin{equation}
\label{r11}
f_{1}=\frac{\left(1-A\right)^{2}}{g^{2}},
\end{equation}
\begin{equation}
\label{r12}
f_{2}=\frac{1}{2g^{2}}\left[g^{2}-\left(1-A\right)^{2}+\frac{\left(1-A-g^{2}\right)^{2}}{1-g^{2}}\right],
\end{equation}
\begin{equation}
\label{r13}
f_{3}=-\frac{1}{2g^{2}}\left[\frac{\left(1-A-g^{2}\right)^{2}}{1-g^{2}}\right].
\end{equation}
The symbol $\Psi$ denotes the digamma function.

We solved numerically equation \eq{r9} assuming the input parameters for the pressure at $130$ GPa. It turns out that equation \eq{r9} allows to reproduce the experimental values of the critical temperature and the isotope coefficient for $A=0.904$ and $B=29.12$ meV. Physically this means the very sharp drop in the electron density of states at and near the Fermi level in the narrow energy range. The obtained result in the natural manner can be associated with the offset of the $\rho\left(\varepsilon\right)$ peak from the Fermi surface.

In conclusion, basing on the experimental data we determined the range of variation of the isotopic coefficient for ${\rm H_{3}S}$ superconductor in the function of the pressure. We showed that the isotope coefficient accepts the anomalously high values in the area of the lower pressures ($\sim 130$ GPa). On the other hand, for the higher pressures ($\sim 200$ GPa), the values of $\alpha$ are lower than those in the BCS theory. The conducted theoretical analysis proved that the low values of the isotope coefficient could be reproduced in the framework of the classical Eliashberg formalism. The anomalously high values of $\alpha$ could be induced by the strong renormalization of the normal state associated with the significant changes of the electron density of states with the change in the pressure. 
Note that the proposed model does not require the non-phonon pairing mechanism.     


\begin{thebibliography}{42}
\expandafter\ifx\csname natexlab\endcsname\relax\def\natexlab#1{#1}\fi
\expandafter\ifx\csname bibnamefont\endcsname\relax
  \def\bibnamefont#1{#1}\fi
\expandafter\ifx\csname bibfnamefont\endcsname\relax
  \def\bibfnamefont#1{#1}\fi
\expandafter\ifx\csname citenamefont\endcsname\relax
  \def\citenamefont#1{#1}\fi
\expandafter\ifx\csname url\endcsname\relax
  \def\url#1{\texttt{#1}}\fi
\expandafter\ifx\csname urlprefix\endcsname\relax\def\urlprefix{URL }\fi
\providecommand{\bibinfo}[2]{#2}
\providecommand{\eprint}[2][]{\url{#2}}

\bibitem[{\citenamefont{Wigner and Huntington}(1935)}]{Wigner1935A}
\bibinfo{author}{\bibfnamefont{E.}~\bibnamefont{Wigner}} \bibnamefont{and}
  \bibinfo{author}{\bibfnamefont{H.~B.} \bibnamefont{Huntington}},
  \bibinfo{journal}{The Journal of Chemical Physics}
  \textbf{\bibinfo{volume}{3}}, \bibinfo{pages}{764} (\bibinfo{year}{1935}).

\bibitem[{\citenamefont{Ashcroft}(1968)}]{Ashcroft1968A}
\bibinfo{author}{\bibfnamefont{N.~W.} \bibnamefont{Ashcroft}},
  \bibinfo{journal}{Physical Review Letters} \textbf{\bibinfo{volume}{21}},
  \bibinfo{pages}{1748} (\bibinfo{year}{1968}).

\bibitem[{\citenamefont{Cudazzo et~al.}(2008)\citenamefont{Cudazzo, Profeta,
  Sanna, Floris, Continenza, Massidda, and Gross}}]{Cudazzo2008A}
\bibinfo{author}{\bibfnamefont{P.}~\bibnamefont{Cudazzo}},
  \bibinfo{author}{\bibfnamefont{G.}~\bibnamefont{Profeta}},
  \bibinfo{author}{\bibfnamefont{A.}~\bibnamefont{Sanna}},
  \bibinfo{author}{\bibfnamefont{A.}~\bibnamefont{Floris}},
  \bibinfo{author}{\bibfnamefont{A.}~\bibnamefont{Continenza}},
  \bibinfo{author}{\bibfnamefont{S.}~\bibnamefont{Massidda}}, \bibnamefont{and}
  \bibinfo{author}{\bibfnamefont{E.~K.~U.} \bibnamefont{Gross}},
  \bibinfo{journal}{Physical Review Letters} \textbf{\bibinfo{volume}{100}},
  \bibinfo{pages}{257001} (\bibinfo{year}{2008}).

\bibitem[{\citenamefont{Szcz{\c{e}}{\'s}niak and
  Jarosik}(2009)}]{Szczesniak2009A}
\bibinfo{author}{\bibfnamefont{R.}~\bibnamefont{Szcz{\c{e}}{\'s}niak}}
  \bibnamefont{and} \bibinfo{author}{\bibfnamefont{M.~W.}
  \bibnamefont{Jarosik}}, \bibinfo{journal}{Solid State Communications}
  \textbf{\bibinfo{volume}{149}}, \bibinfo{pages}{2053} (\bibinfo{year}{2009}).

\bibitem[{\citenamefont{McMahon and Ceperley}(2011)}]{McMahon2011A}
\bibinfo{author}{\bibfnamefont{J.~M.} \bibnamefont{McMahon}} \bibnamefont{and}
  \bibinfo{author}{\bibfnamefont{D.~M.} \bibnamefont{Ceperley}},
  \bibinfo{journal}{Physical Review B} \textbf{\bibinfo{volume}{84}},
  \bibinfo{pages}{144515} (\bibinfo{year}{2011}).

\bibitem[{\citenamefont{Narayana et~al.}(1998)\citenamefont{Narayana, Luo,
  Orloff, and Ruoff}}]{Narayana1998A}
\bibinfo{author}{\bibfnamefont{C.}~\bibnamefont{Narayana}},
  \bibinfo{author}{\bibfnamefont{H.}~\bibnamefont{Luo}},
  \bibinfo{author}{\bibfnamefont{J.}~\bibnamefont{Orloff}}, \bibnamefont{and}
  \bibinfo{author}{\bibfnamefont{A.~L.} \bibnamefont{Ruoff}},
  \bibinfo{journal}{Nature} \textbf{\bibinfo{volume}{393}}, \bibinfo{pages}{46}
  (\bibinfo{year}{1998}).

\bibitem[{\citenamefont{Stadele and Martin}(2000)}]{Stadele2000A}
\bibinfo{author}{\bibfnamefont{M.}~\bibnamefont{Stadele}} \bibnamefont{and}
  \bibinfo{author}{\bibfnamefont{R.~M.} \bibnamefont{Martin}},
  \bibinfo{journal}{Physical Review Letters} \textbf{\bibinfo{volume}{84}},
  \bibinfo{pages}{6070} (\bibinfo{year}{2000}).

\bibitem[{\citenamefont{Ashcroft}(2004)}]{Ashcroft2004A}
\bibinfo{author}{\bibfnamefont{N.~W.} \bibnamefont{Ashcroft}},
  \bibinfo{journal}{Physical Review Letters} \textbf{\bibinfo{volume}{92}},
  \bibinfo{pages}{187002} (\bibinfo{year}{2004}).

\bibitem[{\citenamefont{Kim et~al.}(2010)\citenamefont{Kim, Scheicher, Kang,
  and Ahuja}}]{Kim2010A}
\bibinfo{author}{\bibfnamefont{D.~Y.} \bibnamefont{Kim}},
  \bibinfo{author}{\bibfnamefont{H.~M. R.~H.} \bibnamefont{Scheicher}},
  \bibinfo{author}{\bibfnamefont{T.~W.} \bibnamefont{Kang}}, \bibnamefont{and}
  \bibinfo{author}{\bibfnamefont{R.}~\bibnamefont{Ahuja}},
  \bibinfo{journal}{Proceedings of the National Academy of Sciences}
  \textbf{\bibinfo{volume}{107}}, \bibinfo{pages}{2793} (\bibinfo{year}{2010}).

\bibitem[{\citenamefont{Goncharenko et~al.}(2008)\citenamefont{Goncharenko,
  Eremets, Hanfland, Tse, Amboage, Yao, and Trojan}}]{Goncharenko2008A}
\bibinfo{author}{\bibfnamefont{I.}~\bibnamefont{Goncharenko}},
  \bibinfo{author}{\bibfnamefont{M.~I.} \bibnamefont{Eremets}},
  \bibinfo{author}{\bibfnamefont{M.}~\bibnamefont{Hanfland}},
  \bibinfo{author}{\bibfnamefont{J.~S.} \bibnamefont{Tse}},
  \bibinfo{author}{\bibfnamefont{M.}~\bibnamefont{Amboage}},
  \bibinfo{author}{\bibfnamefont{Y.}~\bibnamefont{Yao}}, \bibnamefont{and}
  \bibinfo{author}{\bibfnamefont{I.~A.} \bibnamefont{Trojan}},
  \bibinfo{journal}{Physical Review Letters} \textbf{\bibinfo{volume}{100}},
  \bibinfo{pages}{045504} (\bibinfo{year}{2008}).

\bibitem[{\citenamefont{Gao et~al.}(2011)\citenamefont{Gao, Wang, Li, Liu, and
  Ma}}]{Gao2011A}
\bibinfo{author}{\bibfnamefont{G.}~\bibnamefont{Gao}},
  \bibinfo{author}{\bibfnamefont{H.}~\bibnamefont{Wang}},
  \bibinfo{author}{\bibfnamefont{Y.}~\bibnamefont{Li}},
  \bibinfo{author}{\bibfnamefont{G.}~\bibnamefont{Liu}}, \bibnamefont{and}
  \bibinfo{author}{\bibfnamefont{Y.}~\bibnamefont{Ma}},
  \bibinfo{journal}{Physical Review B} \textbf{\bibinfo{volume}{84}},
  \bibinfo{pages}{064118} (\bibinfo{year}{2011}).

\bibitem[{\citenamefont{Eremets et~al.}(2008)\citenamefont{Eremets, Trojan,
  Medvedev, Tse, and Yao}}]{Eremets2008A}
\bibinfo{author}{\bibfnamefont{M.~I.} \bibnamefont{Eremets}},
  \bibinfo{author}{\bibfnamefont{I.~A.} \bibnamefont{Trojan}},
  \bibinfo{author}{\bibfnamefont{S.~A.} \bibnamefont{Medvedev}},
  \bibinfo{author}{\bibfnamefont{J.~S.} \bibnamefont{Tse}}, \bibnamefont{and}
  \bibinfo{author}{\bibfnamefont{Y.}~\bibnamefont{Yao}},
  \bibinfo{journal}{Science} \textbf{\bibinfo{volume}{319}},
  \bibinfo{pages}{1506} (\bibinfo{year}{2008}).

\bibitem[{\citenamefont{Martinez-Canales
  et~al.}(2009)\citenamefont{Martinez-Canales, Oganov, Ma, Yan, Lyakhov, and
  Bergara}}]{Martinez2009A}
\bibinfo{author}{\bibfnamefont{M.}~\bibnamefont{Martinez-Canales}},
  \bibinfo{author}{\bibfnamefont{A.~R.} \bibnamefont{Oganov}},
  \bibinfo{author}{\bibfnamefont{Y.}~\bibnamefont{Ma}},
  \bibinfo{author}{\bibfnamefont{Y.}~\bibnamefont{Yan}},
  \bibinfo{author}{\bibfnamefont{A.~O.} \bibnamefont{Lyakhov}},
  \bibnamefont{and} \bibinfo{author}{\bibfnamefont{A.}~\bibnamefont{Bergara}},
  \bibinfo{journal}{Physical Review Letters} \textbf{\bibinfo{volume}{102}},
  \bibinfo{pages}{087005} (\bibinfo{year}{2009}).

\bibitem[{\citenamefont{Chen et~al.}(2008)\citenamefont{Chen, Wang, Struzhkin,
  Mao, Hemley, and Lin}}]{Chen2008A}
\bibinfo{author}{\bibfnamefont{X.~J.} \bibnamefont{Chen}},
  \bibinfo{author}{\bibfnamefont{J.~L.} \bibnamefont{Wang}},
  \bibinfo{author}{\bibfnamefont{V.~V.} \bibnamefont{Struzhkin}},
  \bibinfo{author}{\bibfnamefont{H.}~\bibnamefont{Mao}},
  \bibinfo{author}{\bibfnamefont{R.~J.} \bibnamefont{Hemley}},
  \bibnamefont{and} \bibinfo{author}{\bibfnamefont{H.~Q.} \bibnamefont{Lin}},
  \bibinfo{journal}{Physical Review Letters} \textbf{\bibinfo{volume}{101}},
  \bibinfo{pages}{077002} (\bibinfo{year}{2008}).

\bibitem[{\citenamefont{Tse et~al.}(2007)\citenamefont{Tse, Yao, and
  Tanaka}}]{Tse2007A}
\bibinfo{author}{\bibfnamefont{J.~S.} \bibnamefont{Tse}},
  \bibinfo{author}{\bibfnamefont{Y.}~\bibnamefont{Yao}}, \bibnamefont{and}
  \bibinfo{author}{\bibfnamefont{K.}~\bibnamefont{Tanaka}},
  \bibinfo{journal}{Physical Review Letters} \textbf{\bibinfo{volume}{98}},
  \bibinfo{pages}{117004} (\bibinfo{year}{2007}).

\bibitem[{\citenamefont{Gao et~al.}(2008)\citenamefont{Gao, Oganov, Bergara,
  Martinez-Canales, Cui, Iitaka, Ma, and Zou}}]{Gao2008A}
\bibinfo{author}{\bibfnamefont{G.}~\bibnamefont{Gao}},
  \bibinfo{author}{\bibfnamefont{A.~R.} \bibnamefont{Oganov}},
  \bibinfo{author}{\bibfnamefont{A.}~\bibnamefont{Bergara}},
  \bibinfo{author}{\bibfnamefont{M.}~\bibnamefont{Martinez-Canales}},
  \bibinfo{author}{\bibfnamefont{T.}~\bibnamefont{Cui}},
  \bibinfo{author}{\bibfnamefont{T.}~\bibnamefont{Iitaka}},
  \bibinfo{author}{\bibfnamefont{Y.}~\bibnamefont{Ma}}, \bibnamefont{and}
  \bibinfo{author}{\bibfnamefont{G.}~\bibnamefont{Zou}},
  \bibinfo{journal}{Physical Review Letters} \textbf{\bibinfo{volume}{101}},
  \bibinfo{pages}{107002} (\bibinfo{year}{2008}).

\bibitem[{\citenamefont{Jin et~al.}(2010)\citenamefont{Jin, Meng, He, Ma, Liu,
  Cui, Zou, and Mao}}]{Jin2010A}
\bibinfo{author}{\bibfnamefont{X.}~\bibnamefont{Jin}},
  \bibinfo{author}{\bibfnamefont{X.}~\bibnamefont{Meng}},
  \bibinfo{author}{\bibfnamefont{Z.}~\bibnamefont{He}},
  \bibinfo{author}{\bibfnamefont{Y.}~\bibnamefont{Ma}},
  \bibinfo{author}{\bibfnamefont{B.}~\bibnamefont{Liu}},
  \bibinfo{author}{\bibfnamefont{T.}~\bibnamefont{Cui}},
  \bibinfo{author}{\bibfnamefont{G.}~\bibnamefont{Zou}}, \bibnamefont{and}
  \bibinfo{author}{\bibfnamefont{H.}~\bibnamefont{Mao}},
  \bibinfo{journal}{Proceedings of the National Academy of Sciences}
  \textbf{\bibinfo{volume}{107}}, \bibinfo{pages}{9969} (\bibinfo{year}{2010}).

\bibitem[{\citenamefont{Flores-Livas et~al.}(2012)\citenamefont{Flores-Livas,
  Amsler, Lenosky, Lehtovaara, and Botti}}]{Flores2012A}
\bibinfo{author}{\bibfnamefont{J.~A.} \bibnamefont{Flores-Livas}},
  \bibinfo{author}{\bibfnamefont{M.}~\bibnamefont{Amsler}},
  \bibinfo{author}{\bibfnamefont{T.~J.} \bibnamefont{Lenosky}},
  \bibinfo{author}{\bibfnamefont{L.}~\bibnamefont{Lehtovaara}},
  \bibnamefont{and} \bibinfo{author}{\bibfnamefont{S.}~\bibnamefont{Botti}},
  \bibinfo{journal}{Physical Review Letters} \textbf{\bibinfo{volume}{108}},
  \bibinfo{pages}{117004} (\bibinfo{year}{2012}).

\bibitem[{\citenamefont{Abe and Ashcroft}(2011)}]{Abe2011A}
\bibinfo{author}{\bibfnamefont{K.}~\bibnamefont{Abe}} \bibnamefont{and}
  \bibinfo{author}{\bibfnamefont{N.~W.} \bibnamefont{Ashcroft}},
  \bibinfo{journal}{Physical Review B} \textbf{\bibinfo{volume}{84}},
  \bibinfo{pages}{104118} (\bibinfo{year}{2011}).

\bibitem[{\citenamefont{Wang et~al.}(2012)\citenamefont{Wang, Tse, Tanaka,
  Iitaka, and Ma}}]{Wang2012A}
\bibinfo{author}{\bibfnamefont{H.}~\bibnamefont{Wang}},
  \bibinfo{author}{\bibfnamefont{J.~S.} \bibnamefont{Tse}},
  \bibinfo{author}{\bibfnamefont{K.}~\bibnamefont{Tanaka}},
  \bibinfo{author}{\bibfnamefont{T.}~\bibnamefont{Iitaka}}, \bibnamefont{and}
  \bibinfo{author}{\bibfnamefont{Y.}~\bibnamefont{Ma}},
  \bibinfo{journal}{Proceedings of the National Academy of Sciences}
  \textbf{\bibinfo{volume}{109}}, \bibinfo{pages}{6463} (\bibinfo{year}{2012}).

\bibitem[{\citenamefont{Li et~al.}(2010)\citenamefont{Li, Gao, Xie, Ma, Cui,
  and Zou}}]{Li2010A}
\bibinfo{author}{\bibfnamefont{Y.}~\bibnamefont{Li}},
  \bibinfo{author}{\bibfnamefont{G.}~\bibnamefont{Gao}},
  \bibinfo{author}{\bibfnamefont{Y.}~\bibnamefont{Xie}},
  \bibinfo{author}{\bibfnamefont{Y.}~\bibnamefont{Ma}},
  \bibinfo{author}{\bibfnamefont{T.}~\bibnamefont{Cui}}, \bibnamefont{and}
  \bibinfo{author}{\bibfnamefont{G.}~\bibnamefont{Zou}},
  \bibinfo{journal}{Proceedings of the National Academy of Sciences}
  \textbf{\bibinfo{volume}{107}}, \bibinfo{pages}{15708}
  (\bibinfo{year}{2010}).

\bibitem[{\citenamefont{Zhong et~al.}(2012)\citenamefont{Zhong, Zhang, Chen,
  Li, Zhang, and Lin}}]{Zhong2012A}
\bibinfo{author}{\bibfnamefont{G.}~\bibnamefont{Zhong}},
  \bibinfo{author}{\bibfnamefont{C.}~\bibnamefont{Zhang}},
  \bibinfo{author}{\bibfnamefont{X.}~\bibnamefont{Chen}},
  \bibinfo{author}{\bibfnamefont{Y.}~\bibnamefont{Li}},
  \bibinfo{author}{\bibfnamefont{R.}~\bibnamefont{Zhang}}, \bibnamefont{and}
  \bibinfo{author}{\bibfnamefont{H.}~\bibnamefont{Lin}}, \bibinfo{journal}{The
  Journal of Physical Chemistry C} \textbf{\bibinfo{volume}{116}},
  \bibinfo{pages}{5225} (\bibinfo{year}{2012}).

\bibitem[{\citenamefont{Zhong et~al.}(2013)\citenamefont{Zhong, Zhang, Wu,
  Song, Liu, and Yang}}]{Zhong2013A}
\bibinfo{author}{\bibfnamefont{G.}~\bibnamefont{Zhong}},
  \bibinfo{author}{\bibfnamefont{C.}~\bibnamefont{Zhang}},
  \bibinfo{author}{\bibfnamefont{G.}~\bibnamefont{Wu}},
  \bibinfo{author}{\bibfnamefont{J.}~\bibnamefont{Song}},
  \bibinfo{author}{\bibfnamefont{Z.}~\bibnamefont{Liu}}, \bibnamefont{and}
  \bibinfo{author}{\bibfnamefont{C.}~\bibnamefont{Yang}},
  \bibinfo{journal}{Physica B} \textbf{\bibinfo{volume}{410}},
  \bibinfo{pages}{90} (\bibinfo{year}{2013}).

\bibitem[{\citenamefont{Drozdov et~al.}(2014)\citenamefont{Drozdov, Eremets,
  and Troyan}}]{Drozdov2014A}
\bibinfo{author}{\bibfnamefont{A.~P.} \bibnamefont{Drozdov}},
  \bibinfo{author}{\bibfnamefont{M.~I.} \bibnamefont{Eremets}},
  \bibnamefont{and} \bibinfo{author}{\bibfnamefont{I.~A.}
  \bibnamefont{Troyan}}, \bibinfo{journal}{arxiv:1412.0460}
  (\bibinfo{year}{2014}).

\bibitem[{\citenamefont{Drozdov et~al.}(2015)\citenamefont{Drozdov, Eremets,
  Troyan, Ksenofontov, and Shylin}}]{Drozdov2015A}
\bibinfo{author}{\bibfnamefont{A.~P.} \bibnamefont{Drozdov}},
  \bibinfo{author}{\bibfnamefont{M.~I.} \bibnamefont{Eremets}},
  \bibinfo{author}{\bibfnamefont{I.~A.} \bibnamefont{Troyan}},
  \bibinfo{author}{\bibfnamefont{V.}~\bibnamefont{Ksenofontov}},
  \bibnamefont{and} \bibinfo{author}{\bibfnamefont{S.~I.}
  \bibnamefont{Shylin}}, \bibinfo{journal}{Nature}
  \textbf{\bibinfo{volume}{525}}, \bibinfo{pages}{73} (\bibinfo{year}{2015}).

\bibitem[{\citenamefont{Einaga et~al.}(2015)\citenamefont{Einaga, Sakata,
  Ishikawa, Shimizu, Eremets, Drozdov, Troyan, Hirao, and
  Ohishi}}]{Einaga2015A}
\bibinfo{author}{\bibfnamefont{M.}~\bibnamefont{Einaga}},
  \bibinfo{author}{\bibfnamefont{M.}~\bibnamefont{Sakata}},
  \bibinfo{author}{\bibfnamefont{T.}~\bibnamefont{Ishikawa}},
  \bibinfo{author}{\bibfnamefont{K.}~\bibnamefont{Shimizu}},
  \bibinfo{author}{\bibfnamefont{M.~I.} \bibnamefont{Eremets}},
  \bibinfo{author}{\bibfnamefont{A.~P.} \bibnamefont{Drozdov}},
  \bibinfo{author}{\bibfnamefont{I.~A.} \bibnamefont{Troyan}},
  \bibinfo{author}{\bibfnamefont{N.}~\bibnamefont{Hirao}}, \bibnamefont{and}
  \bibinfo{author}{\bibfnamefont{Y.}~\bibnamefont{Ohishi}},
  \bibinfo{journal}{arXiv:1509.03156}  (\bibinfo{year}{2015}).

\bibitem[{\citenamefont{Duan et~al.}(2014)\citenamefont{Duan, Liu, Tian, Li,
  Huang, Zhao, Yu, Liu, Tian, and Cui}}]{Duan2014A}
\bibinfo{author}{\bibfnamefont{D.}~\bibnamefont{Duan}},
  \bibinfo{author}{\bibfnamefont{Y.}~\bibnamefont{Liu}},
  \bibinfo{author}{\bibfnamefont{F.}~\bibnamefont{Tian}},
  \bibinfo{author}{\bibfnamefont{D.}~\bibnamefont{Li}},
  \bibinfo{author}{\bibfnamefont{X.}~\bibnamefont{Huang}},
  \bibinfo{author}{\bibfnamefont{Z.}~\bibnamefont{Zhao}},
  \bibinfo{author}{\bibfnamefont{H.}~\bibnamefont{Yu}},
  \bibinfo{author}{\bibfnamefont{B.}~\bibnamefont{Liu}},
  \bibinfo{author}{\bibfnamefont{W.}~\bibnamefont{Tian}}, \bibnamefont{and}
  \bibinfo{author}{\bibfnamefont{T.}~\bibnamefont{Cui}},
  \bibinfo{journal}{Sientific reports} \textbf{\bibinfo{volume}{4}},
  \bibinfo{pages}{6968} (\bibinfo{year}{2014}).

\bibitem[{\citenamefont{Li et~al.}(2014)\citenamefont{Li, Hao, Liu, Li, and
  Ma}}]{Li2014A}
\bibinfo{author}{\bibfnamefont{Y.}~\bibnamefont{Li}},
  \bibinfo{author}{\bibfnamefont{J.}~\bibnamefont{Hao}},
  \bibinfo{author}{\bibfnamefont{H.}~\bibnamefont{Liu}},
  \bibinfo{author}{\bibfnamefont{Y.}~\bibnamefont{Li}}, \bibnamefont{and}
  \bibinfo{author}{\bibfnamefont{Y.}~\bibnamefont{Ma}}, \bibinfo{journal}{The
  Journal of Chemical Physics} \textbf{\bibinfo{volume}{140}},
  \bibinfo{pages}{174712} (\bibinfo{year}{2014}).

\bibitem[{\citenamefont{Durajski
  et~al.}(2015{\natexlab{a}})\citenamefont{Durajski, Szcz{\c{e}}{\'s}niak, and
  Li}}]{Durajski2015A}
\bibinfo{author}{\bibfnamefont{A.~P.} \bibnamefont{Durajski}},
  \bibinfo{author}{\bibfnamefont{R.}~\bibnamefont{Szcz{\c{e}}{\'s}niak}},
  \bibnamefont{and} \bibinfo{author}{\bibfnamefont{Y.}~\bibnamefont{Li}},
  \bibinfo{journal}{Physica C} \textbf{\bibinfo{volume}{515}},
  \bibinfo{pages}{1} (\bibinfo{year}{2015}{\natexlab{a}}).

\bibitem[{\citenamefont{Durajski
  et~al.}(2015{\natexlab{b}})\citenamefont{Durajski, Szcz{\c{e}}{\'s}niak, and
  Pietronero}}]{Durajski2015B}
\bibinfo{author}{\bibfnamefont{A.~P.} \bibnamefont{Durajski}},
  \bibinfo{author}{\bibfnamefont{R.}~\bibnamefont{Szcz{\c{e}}{\'s}niak}},
  \bibnamefont{and}
  \bibinfo{author}{\bibfnamefont{L.}~\bibnamefont{Pietronero}},
  \bibinfo{journal}{Annalen der Physik, DOI:10.1002/andp.20150031}
  (\bibinfo{year}{2015}{\natexlab{b}}).

\bibitem[{\citenamefont{Bianconi and
  Jarlborg}(2015{\natexlab{a}})}]{BianconiEPL}
\bibinfo{author}{\bibfnamefont{A.}~\bibnamefont{Bianconi}} \bibnamefont{and}
  \bibinfo{author}{\bibfnamefont{T.}~\bibnamefont{Jarlborg}},
  \bibinfo{journal}{Europhysics Letters} \textbf{\bibinfo{volume}{112}},
  \bibinfo{pages}{37001} (\bibinfo{year}{2015}{\natexlab{a}}).

\bibitem[{\citenamefont{Ortenzi et~al.}(2015)\citenamefont{Ortenzi, Cappelluti,
  and Pietronero}}]{Cappelluti}
\bibinfo{author}{\bibfnamefont{L.}~\bibnamefont{Ortenzi}},
  \bibinfo{author}{\bibfnamefont{E.}~\bibnamefont{Cappelluti}},
  \bibnamefont{and}
  \bibinfo{author}{\bibfnamefont{L.}~\bibnamefont{Pietronero}},
  \bibinfo{journal}{arXiv: 1511.04304}  (\bibinfo{year}{2015}).

\bibitem[{\citenamefont{Bardeen
  et~al.}(1957{\natexlab{a}})\citenamefont{Bardeen, Cooper, and
  Schrieffer}}]{Bardeen1957A}
\bibinfo{author}{\bibfnamefont{J.}~\bibnamefont{Bardeen}},
  \bibinfo{author}{\bibfnamefont{L.~N.} \bibnamefont{Cooper}},
  \bibnamefont{and} \bibinfo{author}{\bibfnamefont{J.~R.}
  \bibnamefont{Schrieffer}}, \bibinfo{journal}{Physical Review}
  \textbf{\bibinfo{volume}{106}}, \bibinfo{pages}{162}
  (\bibinfo{year}{1957}{\natexlab{a}}).

\bibitem[{\citenamefont{Bardeen
  et~al.}(1957{\natexlab{b}})\citenamefont{Bardeen, Cooper, and
  Schrieffer}}]{Bardeen1957B}
\bibinfo{author}{\bibfnamefont{J.}~\bibnamefont{Bardeen}},
  \bibinfo{author}{\bibfnamefont{L.~N.} \bibnamefont{Cooper}},
  \bibnamefont{and} \bibinfo{author}{\bibfnamefont{J.~R.}
  \bibnamefont{Schrieffer}}, \bibinfo{journal}{Physical Review}
  \textbf{\bibinfo{volume}{108}}, \bibinfo{pages}{1175}
  (\bibinfo{year}{1957}{\natexlab{b}}).

\bibitem[{\citenamefont{Eliashberg}(1960)}]{Eliashberg1960A}
\bibinfo{author}{\bibfnamefont{G.~M.} \bibnamefont{Eliashberg}},
  \bibinfo{journal}{Soviet Physics-JETP} \textbf{\bibinfo{volume}{11}},
  \bibinfo{pages}{696} (\bibinfo{year}{1960}).

\bibitem[{\citenamefont{Szcz{\c{e}}{\'s}niak}(2006)}]{Szczesniak2006A}
\bibinfo{author}{\bibfnamefont{R.}~\bibnamefont{Szcz{\c{e}}{\'s}niak}},
  \bibinfo{journal}{Acta Physica Polonica A} \textbf{\bibinfo{volume}{109}},
  \bibinfo{pages}{179} (\bibinfo{year}{2006}).

\bibitem[{\citenamefont{Morel and Anderson}(1962)}]{Morel1962A}
\bibinfo{author}{\bibfnamefont{P.}~\bibnamefont{Morel}} \bibnamefont{and}
  \bibinfo{author}{\bibfnamefont{P.~W.} \bibnamefont{Anderson}},
  \bibinfo{journal}{Physical Review} \textbf{\bibinfo{volume}{125}},
  \bibinfo{pages}{1263} (\bibinfo{year}{1962}).

\bibitem[{\citenamefont{Errea et~al.}(2015)\citenamefont{Errea, Calandra,
  Pickard, Richard, Needs, Li, Liu, Zhang, Ma, and Mauri}}]{Errea2015A}
\bibinfo{author}{\bibfnamefont{I.}~\bibnamefont{Errea}},
  \bibinfo{author}{\bibfnamefont{M.}~\bibnamefont{Calandra}},
  \bibinfo{author}{\bibfnamefont{C.~J.} \bibnamefont{Pickard}},
  \bibinfo{author}{\bibfnamefont{J.~N.} \bibnamefont{Richard}},
  \bibinfo{author}{\bibfnamefont{J.}~\bibnamefont{Needs}},
  \bibinfo{author}{\bibfnamefont{Y.}~\bibnamefont{Li}},
  \bibinfo{author}{\bibfnamefont{H.}~\bibnamefont{Liu}},
  \bibinfo{author}{\bibfnamefont{Y.}~\bibnamefont{Zhang}},
  \bibinfo{author}{\bibfnamefont{Y.}~\bibnamefont{Ma}}, \bibnamefont{and}
  \bibinfo{author}{\bibfnamefont{F.}~\bibnamefont{Mauri}},
  \bibinfo{journal}{Physical Review Letters} \textbf{\bibinfo{volume}{114}},
  \bibinfo{pages}{157004} (\bibinfo{year}{2015}).

\bibitem[{\citenamefont{Hirsch and Marsiglio}(2015)}]{Hirsch2015A}
\bibinfo{author}{\bibfnamefont{J.~E.} \bibnamefont{Hirsch}} \bibnamefont{and}
  \bibinfo{author}{\bibfnamefont{F.}~\bibnamefont{Marsiglio}},
  \bibinfo{journal}{Physica C} \textbf{\bibinfo{volume}{511}},
  \bibinfo{pages}{45} (\bibinfo{year}{2015}).

\bibitem[{\citenamefont{Papaconstantopoulos
  et~al.}(2015)\citenamefont{Papaconstantopoulos, Klein, Mehl, and
  Pickett}}]{Papaconstantopoulos2015A}
\bibinfo{author}{\bibfnamefont{D.~A.} \bibnamefont{Papaconstantopoulos}},
  \bibinfo{author}{\bibfnamefont{B.~M.} \bibnamefont{Klein}},
  \bibinfo{author}{\bibfnamefont{M.~J.} \bibnamefont{Mehl}}, \bibnamefont{and}
  \bibinfo{author}{\bibfnamefont{W.~E.} \bibnamefont{Pickett}},
  \bibinfo{journal}{Physical Review B} \textbf{\bibinfo{volume}{91}},
  \bibinfo{pages}{184511} (\bibinfo{year}{2015}).

\bibitem[{\citenamefont{Bianconi and
  Jarlborg}(2015{\natexlab{b}})}]{Bianconi2015A}
\bibinfo{author}{\bibfnamefont{A.}~\bibnamefont{Bianconi}} \bibnamefont{and}
  \bibinfo{author}{\bibfnamefont{T.}~\bibnamefont{Jarlborg}},
  \bibinfo{journal}{Novel Superconducting Materials}
  \textbf{\bibinfo{volume}{1}}, \bibinfo{pages}{37}
  (\bibinfo{year}{2015}{\natexlab{b}}).

\bibitem[{\citenamefont{Szcz{\c{e}}{\'s}niak
  et~al.}(2001)\citenamefont{Szcz{\c{e}}{\'s}niak, Mierzejewski, and
  Zieli{\'n}ski}}]{Szczesniak2001A}
\bibinfo{author}{\bibfnamefont{R.}~\bibnamefont{Szcz{\c{e}}{\'s}niak}},
  \bibinfo{author}{\bibfnamefont{M.}~\bibnamefont{Mierzejewski}},
  \bibnamefont{and}
  \bibinfo{author}{\bibfnamefont{J.}~\bibnamefont{Zieli{\'n}ski}},
  \bibinfo{journal}{Physica C} \textbf{\bibinfo{volume}{355}},
  \bibinfo{pages}{126} (\bibinfo{year}{2001}).

\end{thebibliography}

\end{document}